# Successive Cancellation Decoding of Polar Codes using Stochastic Computing

Bo Yuan and Keshab K. Parhi, *Fellow, IEEE*
Department of Electrical and Computer Engineering
University of Minnesota, Twin Cities
Minneapolis, MN, USA
{yuan0103, parhi}@umn.edu

*Abstract*— Polar codes have emerged as the most favorable channel codes for their unique capacity-achieving property. To date, numerous works have been reported for efficient design of polar codes decoder. However, these prior efforts focused on design of polar decoders via deterministic computation, while the behavior of *stochastic* polar decoder, which can have potential advantages such as low complexity and strong error-resilience, has not been studied in existing literatures. This paper, for the first time, investigates polar decoding using stochastic logic. Specifically, the commonly-used successive cancellation (SC) algorithm is reformulated into the stochastic form. Several methods that can potentially improve decoding performance are discussed and analyzed. Simulation results show that a stochastic SC decoder can achieve similar error-correcting performance as its deterministic counterpart. This work can pave the way for future hardware design of stochastic polar codes decoders.

## I. INTRODUCTION

As the first capacity-achieving channel codes [1], polar codes have received significant attention from both coding theory and VLSI design communities. Specifically, researchers have focused on design of efficient polar decoders since the decoder is the key component of channel codec. Various polar code decoding algorithms, including successive cancellation (SC), SC list, belief propagation (BP), have been proposed in [1-3] and further optimized in [4-6]. Furthermore the corresponding hardware architectures of these decoding algorithms have been developed and reported in [7-9]. To date, the research on polar codes is the most active field among all the channel codes.

However, it is noticed that all the prior works on polar decoding have been based on deterministic computation. Although deterministic computing scheme has gained great success in past years, it is facing severe challenges in current nanoscale CMOS era that has stringent requirements on power, fault tolerance and speed. Accordingly, stochastic computing, which has inherent advantage on error resilience and low complexity, is expected to offer its unique benefits to address this problem. To date stochastic computing has been applied in various scenarios, such as image processing, control systems and communication systems. In particular, stochastic decoders were investigated for various channel codes, which includes LDPC codes [10], convolutional codes [11], and LDPC convolutional codes (LDPC-CC) [12]. However, no works on stochastic polar decoders have been reported.

This paper presents architectures for polar code decoders using stochastic computing. First, the commonly-used deterministic SC algorithm is reformulated into the stochastic form. Then, different approaches that can potentially improve the decoding performance are analyzed and discussed. Simulation results show that the stochastic polar SC decoder can achieve similar error-correcting performance as its deterministic counterpart. In summary, this work paves the way for future efficient hardware design of stochastic polar code decoders.

The rest of this paper is organized as follows. Section II gives a brief introduction on polar codes and the deterministic SC decoding algorithm. The stochastic SC decoding algorithm is proposed in Section III. Section IV analyzes different approaches that potentially improve decoding performance. The simulation results are also presented in this Section. Section V draws the conclusions.

## II. POLAR CODES AND DETERMINSTIC SC DECODING

### A. Polar Codes

With the use of channel polarization, polar codes were proposed and developed in [1]. In general, the encoding process of $(n, k)$ polar codes consists of two steps. The first step is to construct an intermediate length-$n$ message $u=(u_1, u_2...u_n)$ from a length-$k$ source message. Because the post-decoding reliabilities of bits are polarized based on their positions in the codeword, efficient polar code encoding usually assigns the bits of source message on the most reliable $k$ positions, while forcing the other $n-k$ positions as bit "0". Then, in the second step, $u$ is multiplied with $n$-by-$n$ generator matrix $G$ to obtain transmitted codeword $x=uG$. For the details of polar code encoding, the reader is referred to [1].

### B. Determinstic SC Decoding

At the receiver end, the transmitted codeword $x$ is corrupted to a received codeword $y=(y_1, y_2...y_n)$. With the use of the likelihood ratio (LR) of $y_i$, a deterministic SC decoder performs decoding procedure to recover $u$. An example SC decoding procedure for $n=4$ polar codes is shown in Fig. 1. Here the decoder consists of two basic nodes, namely **f** node and **g** node, respectively. Notice that the operations of these two nodes are also based on deterministic computations that are described in (1) (2). Besides, the number that is associated with each node is the time index that indicates when the node is activated. In this example, at clock cycle 2, 3, 5 and 6, the **f** or **g** node in the last stage (stage-2) sends its calculated LR output to a hard-decision unit (**h** node) for final binary output.

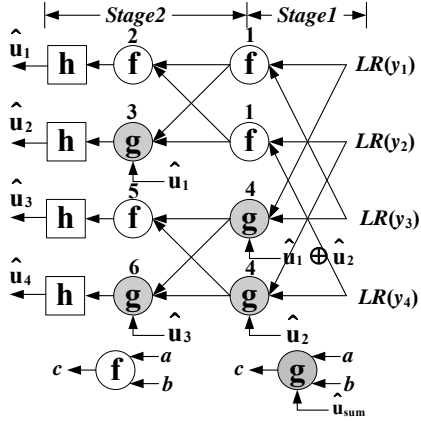

Fig. 1. An example LR-based SC decoding scheme for $n=4$.

$$f(a,b) = \frac{1+ab}{a+b} \tag{1}$$

$$g(a,b,\hat{u}_{sum}) = a^{1-2\hat{u}_{sum}}b. \tag{2}$$

### III. STOCHASTIC SC DECODING

#### A. Stochastic computing

Different from conventional deterministic computation that uses weighted sum to represent values, stochastic computing maps each value to a bit-stream. Here the proportion of number of "1" in the length of the bit-stream corresponds to the represented value. Because the proportion is within [0, 1] range, all the values that are out of this range need to be pre-scaled. Fig. 2(a) shows three different length-10 bit-streams to represent 0.4. It can be seen that the approaches that represent the same value can be different and flexible.

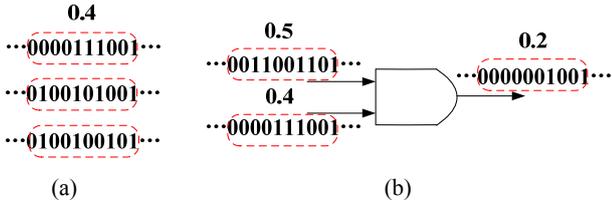

Fig. 2. (a) Represent 0.4 via bit-streams. (b) Stochastic multiplication

Compared with its deterministic counterpart, the stochastic computing has stronger error resilience. That is because the flipping of some bits in the bit-stream does not change the represented value significantly. In addition, the hardware implementation of a stochastic operation has lower complexity. Fig. 2(b) shows the stochastic multiplication that uses only one AND gate. It can be seen that the required hardware resource is far less than the deterministic multiplier that needs hundreds of gates.

#### B. Channel message conversion

In order to design a stochastic SC decoder, we need to first convert the original deterministic channel output in Fig. 1 into stochastic form. Since these channel messages are based on LR form, we can derive the following likelihood information:

$$\Pr(y_i = 1) = \frac{e^{LR(y_i)}}{e^{LR(y_i)}+1}. \tag{3}$$

Notice that $\Pr(y_i=1)$ is within range [0, 1]; hence it can be represented by a bit-stream. As a result, for the stochastic SC decoder, we use the bit-stream that represents $\Pr(y_i=1)$ as the input instead of $LR(y_i)$. This choice is based on the convenient transformation from $\Pr(y_i=1)$ to $\Pr(y_i=0)$ in stochastic computing (only using a NOT gate), and it is also consistent with the case of other stochastic channel decoders [10-13].

Fig. 3 shows the architecture of an input bit-stream generator, which consists of a comparator, a lookup table (LUT) and a pseudo-random number generator.

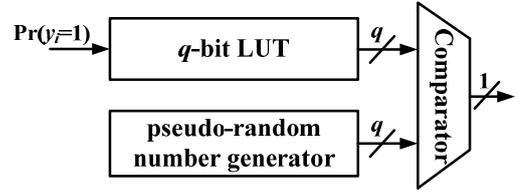

Fig. 3. The architecture of an input bit-stream generator.

#### C. Reformulation of f node

To be compatible with stochastic bit-stream, the original deterministic **f** and **g** nodes in Fig. 1 need to be converted to stochastic forms as well. In this subsection we first consider the reformulation of **f** node.

Recall that the function of **f** node is described in (1), which has two LR-based inputs $a$ and $b$. According to the definition of LR, we have:

$$a = \frac{\Pr(a=0)}{\Pr(a=1)} \qquad b = \frac{\Pr(b=0)}{\Pr(b=1)}.$$

Based on the above denotation, we can re-write (1) as:

$$c = \frac{\Pr(c=0)}{\Pr(c=1)} = f(a,b) = \frac{1+ab}{a+b} = \frac{1+\dfrac{\Pr(a=0)}{\Pr(a=1)}\dfrac{\Pr(b=0)}{\Pr(b=1)}}{\dfrac{\Pr(a=0)}{\Pr(a=1)}+\dfrac{\Pr(b=0)}{\Pr(b=1)}}$$

$$= \frac{\Pr(a=1)\Pr(b=1)+\Pr(a=0)\Pr(b=0)}{\Pr(a=0)\Pr(b=1)+\Pr(a=1)\Pr(b=0)}. \tag{4}$$

From (4) it can be noted that the output of **f** node is the ratio of numerator and denominator, whose sum equals 1. As a result, we have:

$$P_c \triangleq \Pr(c=1) = \Pr(a=0)\Pr(b=1)+\Pr(a=1)\Pr(b=0)$$
$$= P_a(1-P_b)+P_b(1-P_a), \tag{5}$$

where $P_a \triangleq \Pr(a=1)$ and $P_b \triangleq \Pr(b=1)$.

The function of stochastic **f** node is described by (5). Here we use $P_c=\Pr(c=1)$ as the output of **f** node, which is consistent with the choice in Section III-B.

Based on (5), the architecture of stochastic **f** node is developed in Fig. 4, which only needs a XOR gate.

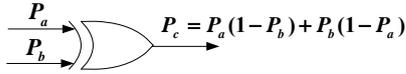

Fig. 4. Architecture of stochastic **f** node.

*D. Reformulation of g node*

Equation (2) describes the function of deterministic **g** node. Again, we need to reformulate this LR-based function to likelihood form. First, consider the case when $\hat{u}_{sum}=0$, then we have:

$$c = \frac{\Pr(c=0)}{\Pr(c=1)} = g(a,b,0) = ab = \frac{\Pr(a=0)}{\Pr(a=1)}\frac{\Pr(b=0)}{\Pr(b=1)}. \quad (6)$$

Notice that the sum of numerator and denominator in (6) is not 1; hence we need to scale it and have Pr($c$=1) as follows:

$$P_c = \Pr(c=1) = \frac{\Pr(a=1)\Pr(b=1)}{\Pr(a=1)\Pr(b=1) + \Pr(a=0)\Pr(b=0)}$$
$$= \frac{P_a P_b}{P_a P_b + (1-P_a)(1-P_b)}, \quad (7)$$

where $P_a$ and $P_b$ are defined in Section III-C.

For the case when $\hat{u}_{sum}=1$, we have:

$$c = \frac{\Pr(c=0)}{\Pr(c=1)} = g(a,b,1) = \frac{b}{a} = \frac{\Pr(a=1)}{\Pr(a=0)}\frac{\Pr(b=0)}{\Pr(b=1)}. \quad (8)$$

Similarly, Pr($c$=1) is derived as follows:

$$P_c = P(c=1) = \frac{P(a=0)P(b=1)}{P(a=0)P(b=1) + P(a=1)P(b=0)}$$
$$= \frac{(1-P_a)P_b}{(1-P_a)P_b + P_a(1-P_b)}. \quad (9)$$

In summary, (7)(9) depict the function of a stochastic **g** node. Accordingly, its hardware architecture is shown in Fig. 5.

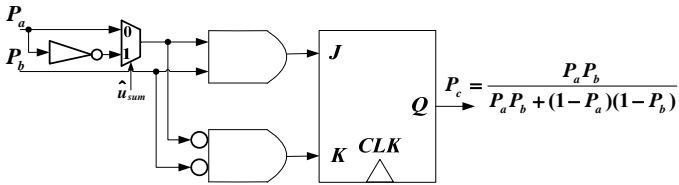

Fig. 5. Architecture of a stochastic **g** node.

## IV. PERFORMANCE DISCUSSION AND SIMULATION RESULTS

In Section III the original deterministic SC decoder has been reformulated to stochastic form. However, due to the approximation in stochastic computing, straightforward use of stochastic **f** and **g** nodes in Fig. 4 and Fig. 5 will cause performance loss. As a result, in this section we analyze several approaches that can potentially improve the decoding performance of stochastic SC decoder.

*A. Channel Message Scaling*

In [10], channel message scaling technique was proposed to improve error-correcting performance of stochastic LDPC decoder. In this approach, $LR(y_i)'=\alpha N_0 LR(y_i)$, instead of the original channel message $LR(y_i)$, is used for generating input bit-streams. Here $N_0$ is the single-sided noise power density. Accordingly, (3) is re-written as:

$$\Pr(y_i=1) \approx \frac{e^{LR(y_i)'}}{e^{LR(y_i)'}+1} = \frac{e^{\alpha N_0 LR(y_i)}}{e^{\alpha N_0 LR(y_i)}+1}. \quad (10)$$

In [10], it was reported that approximation in (10) can partially compensate the performance loss caused by stochastic computation in LDPC decoding. For stochastic SC decoder, a similar phenomenon is observed as well. As shown in Fig. 6, with $\alpha$=0.5, different rate-1/2 stochastic SC decoders achieve significant coding gain over the ones without channel message scaling.

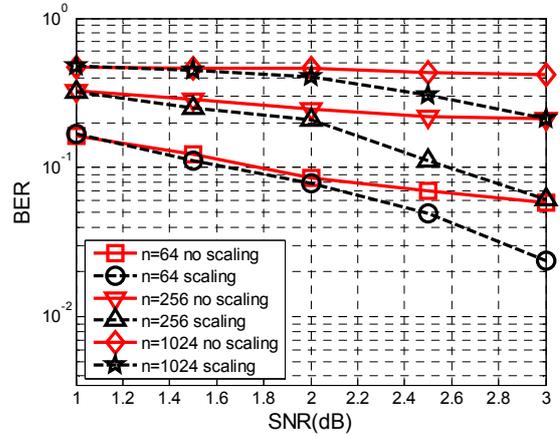

Fig. 6. Simulation results for rate-1/2 stochastic SC decoders with length-128 bit-stream.

*B. Use of Edge Memory*

Another widely adopted technique in stochastic channel decoding is the use of edge memory, which provides extra randomness to avoid "hold" state. However, simulation results show that edge memory in stochastic polar SC decoding is not as efficient as in stochastic LDPC decoding [10]. We find that even when 64-bit edge memories are equipped, no obvious improvement on performance is observed. This is potentially due to the non-iterative procedure of SC algorithm that alleviates the influence of "hold" state.

*C. Increasing length of bit-stream*

In general, the accuracy of stochastic computing is improved with the increase of length of bit-stream. This is due to the improved precision for representation scheme. For the scheme that uses length-$2^s$ bit-stream, the precision is $1/2^s$.

In [10], length-128 bit-stream was used for stochastic LDPC decoding. However, as shown in Fig. 7, a longer bit-stream is required to overcome precision loss in stochastic SC decoding. A possible reason is that polar codes can be approximately viewed as "high-density parity-codes (HDPC)", which has more severe error propagation than LDPC codes. As a result, in the proposed stochastic SC decoding the length of bit-stream is selected as 1024.

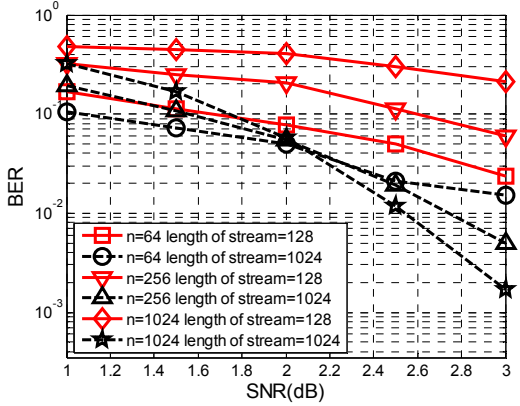

Fig. 7. Simulation results for rate-1/2 stochastic SC decoders with channel message scaling ($\alpha$=0.5).

*D. Rerandomizing bit-stream*

In stochastic computing system, the randomness of generated bit-stream is gradually lost. In that case, re-randomizing the bit-stream is needed. As seen in Fig. 8, with the re-randomizing technique, only negligible performance loss is observed for different rate-1/2 stochastic SC decoders as compared to the conventional deterministic decoders.

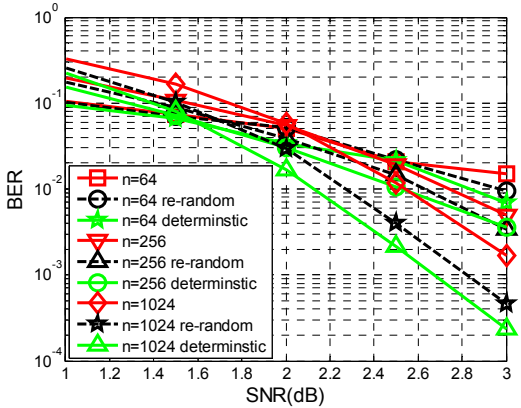

Fig. 8. Simulation results for rate-1/2 stochastic SC decoders with channel message scaling ($\alpha$=0.5) and length-1024 bit-stream.

*E. Discussion & Future work*

According to the simulation results in Fig. 7~Fig. 9, it is seen that stochastic polar SC decoders suffer more severe performance degradation than stochastic LDPC decoders. Longer bit-stream and extra re-randomization, which are not needed in LDPC decoding [10], are mandatory for stochastic polar decoding. These two operations cause longer decoding latency that is the primary drawback for stochastic computing. A potential solution is to use multiple shorter bit-streams for parallel processing. Although this strategy would cause an increase in area, the overall benefit in hardware efficiency can still be significant since the hardware complexity of stochastic computing is much less than its deterministic counterpart.

Future work will be directed towards the approaches that reduce overall latency.

An interesting phenomenon discovered in this paper is the similarity between stochastic polar decoding and LDPC decoding (or more generally, the iterative decoding [13]). It can be seen that architectures of stochastic **f** node in Fig. 4 and **g** node in Fig. 5 are very similar to the stochastic check node and equality node in [10][13]. It is believed that this similarity is due to the inherent property of the SC algorithm that it is a special version of belief propagation (BP) algorithm [3]. Another part of future work will be directed towards the design of stochastic polar BP decoders.

V. CONCLUSION

This paper investigates the performance of stochastic SC polar code decoder. Various potential approaches that can improve decoding performance are analyzed and discussed. It is shown that the stochastic SC decoder achieves similar error-correcting performance to its deterministic counterpart. This work paves the way for future VLSI design of stochastic polar decoders.

REFERENCES

[1] E. Arikan, "Channel polarization: A method for constructing capacity-achieving codes for symmetric binary-input memoryless channels," *IEEE Trans. Inf. Theory*, vol. 55, no. 7, pp. 3051-3073, 2009.

[2] I. Tal and A. Vardy, "List decoding of polar codes," arXiv:1206.0050, May 2012

[3] E. Arıkan, "Polar codes: A pipelined implementation," in *Proc. 4th Int. Symp. Broad. Commun. (ISBC)*, pp. 11-14, July 2010

[4] K. Niu and K. Chen, "Stack decoding of polar codes," *Elect. Lett.*, vol. 48, no. 12, pp. 695-696, 2012.

[5] B. Yuan and K. K. Parhi, "Architecture optimizations for BP polar decoders," in *Proc. IEEE International Conference on Acoustics, Speech and Signal Processing (ICASSP)*, pp. 2654-2658, May 2013.

[6] A. Alamdar-Yazdi and F. R. Kschischang, "A simplified successive-cancellation decoder for polar codes," *IEEE Commun. Lett.*, vol. 15, no. 12, pp. 1378-1380, 2011.

[7] B. Yuan and K.K. Parhi, "Low-Latency successive-cancellation polar decoder architectures using 2-bit decoding," *IEEE Trans. Circuits and Systems-I: Regular Papers*, vol. 61, no. 4, pp. 1241-1254, Apr. 2014.

[8] B. Yuan and K.K. Parhi, "Early Stopping Criteria for Low-Power Low-Latency Belief-Propagation Polar Code Decoders," accepted by *IEEE Trans. Signal Processing*

[9] B. Yuan and K.K. Parhi, "Low-Latency Successive-Cancellation List Decoders for Polar Codes with Multi-bit Decision," accepted by *IEEE Trans. VLSI Systems*, DOI 10.1109/TVLSI.2014.2359793.

[10] S. Sharifi Tehrani, S. Mannor, and W. J. Gross, "Fully parallel stochastic LDPC decoders," *IEEE Trans. on Signal Processing*, vol. 56, no. 11, pp. 5692–5703, Nov.2008.

[11] T-H. Chen and J.P. Hayes, "Design of Stochastic Viterbi Decoders for Convolutional Codes," in *Proc. of IEEE Int Design and Diagnostics of Electronic Ciruits. & Systems. (DDECS)*, pp. 66-71, April 2013.

[12] X-R Lee, C-L Chen, H-C Chang and C-Y Lee, "Stochastic decoding for LDPC convolutional codes," in *Proc. IEEE Int. Symp. Circuits Syst.(ISCAS)*, pp. 2621-2624, May 2012.

[13] V.C. Gaudet and A.C.Rapley, "Iterative decoding using stochstic computation," IET Elctronics Letters, vol. 39, no. 3, pp 299-301, Feb. 2003.